\documentclass[10pt, conference]{IEEEtran}
\IEEEoverridecommandlockouts
\usepackage{booktabs}   
\usepackage{subcaption} 

\usepackage{mathptmx}
\usepackage{microtype}

\usepackage{amsmath}
\usepackage{newtxmath}

\def\BibTeX{{\rm B\kern-.05em{\sc i\kern-.025em b}\kern-.08em
    T\kern-.1667em\lower.7ex\hbox{E}\kern-.125emX}}

\DeclareSymbolFont{largesymbolsCM}{OMX}{cmex}{m}{n}

\let\sum\relax
\DeclareMathSymbol{\sum}{\mathop}{largesymbolsCM}{"50}

\DeclareSymbolFont{tienlargesymbolsCM}{OMX}{cmex}{m}{n}

\let\prod\relax
\DeclareMathSymbol{\prod}{\mathop}{tienlargesymbolsCM}{"51}

\usepackage{balance}

\usepackage{ulem}

\usepackage[table]{xcolor}

\usepackage{listings}
\usepackage{epsf}
\usepackage{ctable}
\usepackage{graphicx}
\usepackage{url}
\usepackage{float}
\usepackage{multicol}
\restylefloat{figure}
\usepackage{comment}
\usepackage{paralist}
\usepackage{cite}

\usepackage{xspace}

\newcommand{\eg}{\hbox{\emph{e.g.,}}\xspace}
\newcommand{\ie}{\hbox{\emph{i.e.,}}\xspace}

\newcommand{\model}{\textsc{RUBY}\xspace}

\usepackage{multirow}
\usepackage{textcomp}

\usepackage[T1]{fontenc}
\usepackage{array}

\usepackage{fancyhdr}
\usepackage[yyyymmdd,hhmmss]{datetime}
\usepackage{algorithm}
\usepackage[noend]{algpseudocode}

\newcommand{\code}[1]{{\small\textsf{#1}}}

\definecolor{deletedline}{RGB}{255,224,224}
\definecolor{addedline}{RGB}{224,255,224}
\definecolor{modifiedline}{RGB}{231,231,152}

\ifCLASSINFOpdf
\else
\fi
\begin{document}

\title{Does BLEU Score Work for Code Migration?}

\author{\IEEEauthorblockN{Ngoc Tran\IEEEauthorrefmark{1},
Hieu Tran\IEEEauthorrefmark{1},
Son Nguyen\IEEEauthorrefmark{1},
Hoan Nguyen\IEEEauthorrefmark{2}, and
Tien N. Nguyen\IEEEauthorrefmark{1}}
  \IEEEauthorblockA{\IEEEauthorrefmark{1}Computer Science Department, The University of Texas at Dallas, USA,\\ Email: \{nmt140230,trunghieu.tran,sonnguyen,tien.n.nguyen\}@utdallas.edu}
  \IEEEauthorblockA{\IEEEauthorrefmark{2}Computer Science Department,
Iowa State University, USA, Email: hoan@iastate.edu}}

\maketitle

\begin{abstract}
Statistical machine translation (SMT) is a fast-growing sub-field of
computational linguistics. Until now, the most popular automatic metric
to measure the quality of SMT is BiLingual Evaluation Understudy
(BLEU) score. Lately, SMT along with the BLEU metric has been applied
to a Software Engineering task named {\em code migration}. 
(In)Validating the use of BLEU score could advance the research and
development of SMT-based code migration tools. Unfortunately, there is
no study to approve or disapprove the use of BLEU score for source
code.
In this paper, we conducted an empirical study on BLEU score to
(in)validate its suitability for the code migration task due to its
inability to reflect the semantics of source code. In our work, we use
human judgment as the ground truth to measure the semantic correctness of
the migrated code. Our empirical study demonstrates that BLEU
does not reflect translation quality due to its weak correlation with
the semantic correctness of translated code. We provided
counter-examples to show that BLEU is ineffective in comparing the
translation quality between SMT-based models. Due to BLEU's
ineffectiveness for code migration task, we propose an alternative
metric {\model}, which considers lexical, syntactical, and semantic
representations of source code. We verified that {\model} achieves a
higher correlation coefficient with the semantic correctness of migrated
code, 0.775 in comparison with 0.583 of BLEU score. We also confirmed
the effectiveness of {\model} in reflecting the changes in translation
quality of SMT-based translation models. With its advantages, {\model}
can be used to evaluate SMT-based code migration~models.

\end{abstract}


%
\IEEEpeerreviewmaketitle

\section{Introduction}
\label{sec:intro}

Statistical Machine Translation (SMT)~\cite{smtbook} is a
natural~language processing (NLP) approach that uses statistical
learning to derive the translation ``rules'' from a training data and
applies the trained model to translate a sequence from the source
language to the target one.
SMT models are trained from a corpus of corresponding texts in two
languages. SMT has been successful in translating natural-language
texts.  Google Translate~\cite{googletranslate} is an SMT-based
tool that support text translation of over 100
languages. Microsoft Translator~\cite{mstranslator} also supports
instant translation for more than 60 languages.

The statistical machine translation community relies on the {\em
  BiLingual Evaluation Understudy} (BLEU) metric for the purpose of
evaluating SMT models. {\em BLEU metric}, also called {\em
  BLEU~score}, measures translation quality by the accuracy of
translating text phrases to the reference with various phrases'
lengths. BLEU score was shown to be highly correlated with human judgments
on the translated texts from natural-language SMT
tools~\cite{Papineni2002}.
%
BLEU remains a popular automated and inexpensive
metric to evaluate the quality of translation models.


In recent years, several researchers in Software Engineering (SE) and
Programming Languages have been exploring the NLP techniques and
models to build automated SE tools. SMT has been directly used or
adapted to be used to translate/migrate source code in different
programming
languages~\cite{fse13-nier,icse14-demo,karaivanov14,ase15,icsme16}. The
problem is called {\em code migration}. In the modern world of
computing, code migration is important. Software vendors often develop
a product for multiple operating platforms in different languages. For
example, a mobile app could be developed for iOS (in Objective-C),
Android (in Java), and Windows Phone (in C\texttt{\#}).
Thus, there is an increasing need for migration of source code from
one programming language to another.

Unlike natural-language texts, source code follows syntactic rules and
has well-defined semantics with respect to programming languages. A
natural question is {\em how effective BLEU score is in evaluating the
  results of migrated source code in language migration}. The answer
to this question is important because if it does, we could~establish
an automated metric to evaluate the quality of SMT-based code
migration tools. Unfortunately, there has not yet any empirical
evidence to either validate or invalidate the effectiveness of BLEU
score in applying to source code in language migration.

Because BLEU score measures the lexical phrase-to-phrase
translation while source code has well syntax and dependencies, we
hypothesized that {\em BLEU score is not effective in evaluating
  migrated source code}.
We proved this hypothesis by contradiction. First, we assumed that
{\em BLEU score is effective in evaluating migrated source code}.
With respect to the use of BLEU for the translation results by a
single model or its use to compare the results across models, the
above assumption leads to {\em two necessary conditions}
(1) BLEU reflects well the semantic accuracy of migrated source code
with regard to the original code when they are translated by a model,
and (2) the quality comparison between translation models can be drawn
by using the BLEU metric.
%
%
%
We then conducted experiments to provide counterexamples to
empirically invalidate those two conditions. We choose the code
migration task because it is a popular SE task that applies SMT.

For the first condition, we chose the migration models that focus on
phrase translation, for example, lpSMT~\cite{fse13-nier} that adapts a
phrase-to-phrase translation model~\cite{phrasal10}. This type of
models produces migrated code with high lexical accuracy, \ie high
correctness for sequences of code tokens. However, several tokens or
sequences of tokens are placed in incorrect locations.  This results
in a higher BLEU score but with a lower semantic accuracy in migrated
code. We also chose the migration models that focus on
structures. Specifically, we picked mppSMT~\cite{ase15} whose results have
high semantic accuracy but a wide range of BLEU
scores. In this case, we aimed to show that BLEU has weak correlation
with semantic accuracy of the translated code.
For the second condition, adapting Callison-Burch {\em et
  al.}~\cite{Callison}'s method, we constructed an artificial model,
p-mppSMT, from mppSMT~\cite{ase15} that produces the migrated code
with the same BLEU score as the result from mppSMT. However, a result
from p-mppSMT for a given method contains a minor lexical difference
from the one from mppSMT, thus, creating as a consequence 
the considerable differences in program semantics. That is, p-mppSMT
produces migrated code that has the same BLEU score but is
functionally different.
Finally, since the~two necessary conditions are false,
the assumption statement is~false, thus, BLEU is ineffective in
evaluating migrated source code.

In our experiment, we used a dataset of 34,209 pairs of methods in 9
projects that were manually migrated from Java to C\texttt{\#} by
developers. The dataset was used in evaluating existing code migration
models~\cite{ase15}. We ran those above SMT-based migration models on
those methods. We then manually investigated 375 randomly-selected
pairs of methods for each model. For each pair of the manually
migrated method and the automatically migrated one from a tool, we
assigned a semantic accuracy score and computed the BLEU score. The
semantic accuracy score was given by a developer after examining the
original code in Java, the manually migrated code in C\texttt{\#} and
the code in C\texttt{\#} migrated by a tool. In total, 2,250
translated results were manually checked and assigned with semantic
scores in about 45 hours. For the necessary condition (1), we then
computed the correlation between the BLEU scores and semantic scores
of all the pairs.  Our result shows that the BLEU metric has a weak
correlation with the semantic accuracy of the migrated code. For the
necessary condition~(2), we compared the translated results of the
same set of 375 methods for mppSMT~\cite{ase15} and its artificial
variant, p-mppSMT.
%
We showed that the equivalence of BLEU scores of models does not indicate
the equivalence in the quality of their translation results.
From those results, we concluded that BLEU is not an effective~metric
in evaluating 
SMT-based migration tools.



In this work, we also introduce, {\model}, a novel metric to evaluate
the results of the SMT-based code migration tools. The intuition is
that the metric that measures the results in the higher abstraction
level, the better metric for reflecting the semantics
accuracy. However, in code migration, the translated code might miss
information that is required to construct the higher level
representations. Therefore, we propose a {\em multi-level metric to
evaluate the quality of translated code, that integrates the
measurement in three representation levels of source code in a
multi-layer and increasing manner from lexeme (text), abstract syntax
tree (AST), to program dependence graph (PDG)}. For a translated
result, {\model} score is the similarity score between the migrated code
and the expected result in the highest representation level that can
be constructed for both versions. Furthermore,
we also conducted two experiments to evaluate {\model} on the abilities to
reflect the semantic accuracy and the comparison in translation quality
between models. Our results show that the new metric {\model} is
highly correlated to the semantic accuracy of the resulting migrated
code. The average correlation coefficients between {\model} and
semantic accuracy is 0.775, in comparison to 0.583 of BLEU score.  Our
consistent results of {\model} with semantic scores also demonstrate
the effectiveness of {\model} in comparing SMT-based code migration
models.


In summary, in this paper, our main contributions include:

\begin{itemize}
	\item An empirical evidence to show that BLEU metric does not only reflect
well the semantic accuracy of the migrated code for SMT-based
migration tools, but is also not effective in comparing translation
quality of models.

	\item {\model}, a novel metric to evaluate the results of the SMT-based
code migration tools, that integrates the scores at the lexical,
syntactical, and semantic levels in source code.

	\item An extensive empirical study to evaluate {\model} on multiple SMT-based code migration systems.
\end{itemize}

\section{Background}
\label{sec:background}

\subsection{Statistical Machine Translation}

Machine Translation aims to translate texts or speech from a language
to another. Statistical Machine Translation (SMT) is a machine
translation paradigm that uses statistical models to learn to derive
the translation ``rules'' from a training corpus in order to translate
a sequence from the source language ($L_S$) to the target language
($L_T$). The text in the $L_S$ is tokenized into a sequence \textit{s}
of words. The model searches the most relevant target sequence
\textit{t} with respect to \textit{s}. Formally, the model searches
for the target sequence \textit{t} that has the maximum probability:
$$ P\left(t \mid s \right) = \frac{P\left(s \mid t\right) \, P\left(t\right)}{P\left(s\right)} $$

To do that, it utilizes two models: 1) the language model, and 2)
the translation model. The language model learns from the monolingual
corpus of $L_T$ to derive the possibility of feasible sequences in it
($P\left(t\right)$: how likely sequence \textit{t} occurs in
$L_T$). On the other hand, the translation model computes the
likelihood $P\left(s \mid t\right)$ of the mapping between $s$ and
$t$. The mapping is calculated by analyzing the bilingual dual corpus
to learn the alignment between the words or sequences of
two~languages.

\subsection{SMT-based Code Migration}

Traditionally, SMT is widely used in translating natural-language
texts~\cite{smtbook}. With the success of SMT, several researchers
have adapted it to use in programming languages to migrate source code
from one language to another (called {\em code migration} or {\em
  language migration}). lpSMT~\cite{fse13-nier} is a model that
directly applies Phrasal, a phrase-based SMT tool, to migrate Java
code to C\texttt{\#}. Source code is treated as a sequence of code
tokens and a Java code fragment is migrated into a fragment in
C\texttt{\#}. Despite that the migrated code is textually similar to
the manually migrated code, the percentage of migrated methods that
are semantically incorrect is high (65.5\%).
Karaivanov {\em et al.}~\cite{karaivanov14} also follow phrase-based SMT to
migrate C\texttt{\#} to Java. They use prefix grammar to consider only
the phrases that are potentially the beginning of some syntactic
units.

Nguyen {\em et al.}~\cite{ase15} developed mppSMT by using a
divide-and-conquer approach~\cite{sudoh15} with syntax-directed
translation. mppSMT constructs from the code the sequence of
annotations for code token types and data types. mppSMT then uses
phrase-based SMT three times on three sequences built from source
code: lexemes, syntactic and type annotations. It integrates the
resulting translated code at the lexical level for all syntactic units
into a larger code. The type annotations help with the translation of
data types and APIs.
%
codeSMT~\cite{icsme16} improves over lpSMT by using well-defined
semantics in programming languages to build a context to guide SMT
translation process. It integrates five types of contextual features
related to semantic relations among code tokens including
occurrence association among elements, data and control
dependencies among program entities, visibility constraints of
entities, and the consistency in declarations and accesses of
variables, fields, and methods.




\subsection{BLEU Metric}

Manual evaluation on SMT's results is time consuming and expensive for
frequent developing tasks of SMT models~\cite{Papineni2002}.
(\underline{B}i\underline{L}ingual \underline{E}valuation
\underline{U}nderstudy)~\cite{Papineni2002} (BLEU) is an automated
metric to replace human efforts in evaluating machine translation
quality.
BLEU uses the modified form of $n$-grams precision and length
difference penalty to evaluate the quality of text generated by SMT
compared to referenced one.
BLEU measures translation quality by the accuracy of translating
$n$-grams to $n$-grams with various values of $n$ (phrases to
phrases):
\[BLEU = BP.{e^{\frac{1}{n}(\log {P_1} + ... + \log {P_n})}}\]
where $BP$ is the {\em brevity penalty value}, which equals to 1 if
the total length (\ie the number of words) of the resulting sentences
is longer than that of the {\em reference sentences} (\ie the correct
sentences in the oracle). Otherwise, it equals to the ratio between
those two lengths. $P_i$ is the metric for the overlapping between
the bag of $i$-grams (repeating items are allowed) appearing in the
resulting sentences and that of $i$-grams appearing in the reference
sentences. Specifically, if $S^{i}_{ref}$ and $S^{i}_{trans}$ are the
bags of $i$-grams appearing in the reference code and in the
translation code respectively, $P_i$ = |$S^{i}_{ref}$ $\cap$
$S^{i}_{trans}$| / |$S^{i}_{trans}$|. The value of \code{BLEU} is
between 0--1. The higher it is, the higher the n-grams precision.

Since $P_i$ represents the accuracy in translating phrases
with $i$ consecutive words, the higher the value of $i$ is used, the
better \code{BLEU} measures translation quality. For example, assume
that a translation \code{Tr} has a high $P_1$ value but a
low~$P_2$. That is, \code{Tr} has high word-to-word accuracy but low
accuracy in translating 2-grams to 2-grams (\eg the word order might
not be respected in the result). Thus, using both $P_1$ and $P_2$ will
measure \code{Tr} better than using only $P_1$. If
translation~sen\-tences are shorter, \code{BP} is smaller and
\code{BLEU} is smaller. If they are too long and more incorrect words
occur, $P_i$ values are smaller, thus, \code{BLEU} is smaller. $P_i$s
are often computed for $i$=1--4.

\section{Hypothesis and Methodology}


BLEU has been widely used in evaluating the translation quality of SMT
models in NLP. It has been empirically~validated to be correlated with
human judgments for the translation quality in natural-language
texts~\cite{Papineni2002}.
%
%
However, using BLEU metric to evaluate
the SMT-based code migration models would raise two issues. First,
programming languages
have well-defined syntaxes and program dependencies. Source code has
strict syntactic structure while natural-language texts are less
strict in that aspect.
%
%
Second, there is a gap between the purpose of BLEU and the task of
evaluating migrated code. BLEU measures the lexical precision of
translating results. However, when evaluating translated source code,
it is more important to consider the semantics/functionality of
the generated code.
The closer the semantics/functionality between the translated~code and
the reference code, the better translation quality is.
%

Due to those two intuitive reasons, we have the hypothesis that {\em
  BLEU score does not measure well the quality of translated results
  that is estimated based on the semantic/functionality similarity
  between the migrated source code and the reference one}. We prove
this by contradiction. First, we~assume that {\em BLUE
  measure well the quality of translated code}. From that assumption
statement, we derive two {\em two necessary conditions}:


{\bf Cond\_1 [Semantic Similarity].} BLEU score reflects well the
semantic similarity between the translated source code and the
reference code in the ground truth.

{\bf Cond\_2 [Model Comparison].} BLEU is effective in comparing the
translation quality of SMT-based migration models.





\subsection{Counter-Examples and Model Selections}

 

We conducted experiments to provide counterexamples to empirically
invalidate the two above conditions. 

If BLEU is a good metric for language migration, it should be
independent of SMT-based migration models and programming
languages. Therefore, choosing programming languages does not affect
our assessment for BLEU. In this work, we choose the language
migration task from Java to C\texttt{\#}, which is considered as one
of the most popular tasks in language migration.

%
%

To disprove {\bf Cond\_1}, we provide counterexamples to show that
BLEU score is not strongly correlated to the semantic accuracy of
migrated code. That is, for a specific translation~model, the higher
BLEU score does not reflect the higher semantic accuracy of translated
code, and the higher semantic accuracy results are not necessarily
reflected by higher BLEU scores.
%
%
%

\subsubsection{{\bf Higher BLEU score does not necessarily reflect higher semantic accuracy}}

We chose the migration models that focus on phrase translation, for
example, lpSMT~\cite{fse13-nier,karaivanov14} that adapts
phrase-to-phrase translation models~\cite{phrasal10}. This type of
models produces migrated code with high lexical accuracy, \ie high
correctness on the sequences of code tokens. However, several tokens
or sequences of tokens are placed in incorrect locations.  This
results in a higher BLEU score but with lower semantic accuracy in
migrated code.

\begin{figure}[t]
\centering
\begin{lstlisting}[basicstyle=\scriptsize\sffamily, stepnumber=1, numbers=left, language=Java, aboveskip=1pt,  belowskip=1pt, numbersep=-5pt]
		// Java code: ClientQueryResult.java
		public ClientQueryResult(Transaction ta, int initialSize){
			super(ta, initialSize);
		}

		// C# code: ClientQueryResult.cs
		public ClientQueryResult(Transaction ta, int initialSize) : base(ta, initialSize) {}

		// Translated by lpSMT:
		public ClientQueryResult(Transaction ta, int initialSize) : base(ta {, initialSize) ; }
\end{lstlisting}
\caption{lpSMT Example~\cite{fse13-nier}}
\label{fig:issueexample2}
\end{figure}


Let us take an example of the resulting migrated code from the tool
lpSMT~\cite{fse13-nier} ({\em Call to Parent Class' Constructor
  \code{super(...)}}). Fig.~\ref{fig:issueexample2} shows an example
of the lpSMT model translating a call to the constructor of a parent
class via \code{super} (lines 2--4). In Java, a call to \code{super} is
made inside the method's body. In contrast, in C\texttt{\#}, a call to the
constructor is made via \code{base} and occurs in the method
signature, \ie prior to the method's body as in
\code{base(ta,initialSize) {}} (line 7). However, in the translation
version, this call was broken into two pieces: one in the method
signature \code{'base(ta'} and one in the method~body\\ \code{`, initialSize);'} (line 10). Thus, the translation code is
syntactically incorrect. In this case, lpSMT translated based on the
lexemes of tokens in the method signature and the body, however, does
not consider the entire syntactic unit of a constructor call to the
parent class in \code{super} for translation. For this example, the
BLEU score with $n$=4 is 0.673. However, the translated code is not even
compiled due to that syntactic error. The effort to find the
misplacement and to correct it is not trivial.

\subsubsection{{\bf Higher semantic accuracy is not necessarily reflected via higher BLEU score}}
%
%
To validate this, we chose the~migration model
mppSMT~\cite{ase15},~which has high semantic accuracy but probably
with a wide range of BLEU~scores.  The key idea behind this choice is
that mppSMT uses syntax-directed translation, \ie it translates the
code within syntactic structures first, and then aggregates the
translated code for all structures to form the final migrated code. To
improve semantic accuracy, mppSMT integrates during the translation
process the mappings of types and API usages between two
languages. This strategy has been efficient in achieving higher
syntactic and semantic accuracy than other existing
techniques~\cite{ase15}. An important characteristic of mppSMT is that
there is a considerable percentage of translated code that is
semantically correct, however, are different from the manual-migrated
code. Specifically, those correct code involves 1) code with different
local variables' names from a reference method, but all variables are
consistently renamed; 2) code with namespaces being added/deleted
to/from a type (\eg \code{new P.A()} vs \code{new A()}); 3) code with
`\code{this}' being added/deleted to/from a field or method; and 4)
code with different uses of syntactic units and APIs having the same
meaning, \eg field accesses versus getters or array accesses versus
method calls. These variations are all correct, however, the BLEU
scores among the pairs of the reference and translated code are quite
different.

\begin{table}[t]
       \small
	\tabcolsep 3pt
  \centering
  \caption{Parallel Corpus}
    \begin{tabular}{l|rrr|rrr|r}
    \toprule
    Project & \multicolumn{3}{c}{Java} \vline& \multicolumn{3}{c}{C\texttt{\#}} \vline& M.Meth \\
    \hline
		& Ver   &File 		&Meth 	&Ver 			&File 		&Meth 		&  \\
		\hline					
    Antlr~\cite{Antlr} 		   & 3.5.0 & 226   	& 3,303  & 3.5.0 	& 223   	& 2,718  	& 1,380 \\
    db4o~\cite{db4o}  		   & 7.2   & 1,771 	& 11,379 & 7.2   	& 1,302  	& 10,930 	& 8,377 \\
    fpml~\cite{fpml}  		   & 1.7   & 138   	& 1,347  & 1.7   	& 140   	& 1,342  	& 506 \\
    Itext~\cite{Itext} 		   & 5.3.5 & 500   	& 6,185  & 5.3.5 	& 462   	& 3,592  	& 2,979 \\
    JGit~\cite{JGit}  		   & 2.3   & 1,008 	& 9,411  & 2.3   	& 1,078  	& 9,494  	& 6,010 \\
    JTS~\cite{JTS}   		   & 1.13  & 449   	& 3,673  & 1.13  	& 422   	& 2,812  	& 2,010 \\
    Lucene~\cite{Lucene}  	   & 2.4.0 & 526   	& 5,007  & 2.4.0 	& 540   	& 6,331  	& 4,515 \\
    Neodatis~\cite{Neodatis}       & 1.9.6 & 950   	& 6,516  & 1.9b-6       & 946           & 7,438 	& 4,399 \\
    POI~\cite{POI}   		   & 3.8.0 & 880        & 8,646  & 1.2.5 	& 962   	& 5,912  	& 4,452 \\
    \bottomrule
    \end{tabular}%
  \label{tab:systems}%
\end{table}%


%

To disprove {\bf Cond\_2}, we provide a counterexample to empirically
show the ineffectiveness of BLEU in comparing the SMT-based migration
models. Specifically, we will show that {\em the equivalence of BLEU
  scores of the results translated by two models is not consistent
  with the equivalence in the translation quality of these models}.

To provide the counterexample, we construct a {\em permuted model,
  p-mppSMT}, from mppSMT, that applies the same principle as in
Callison-Burch {\em et al.}~\cite{Callison}. Their idea is that BLEU
{\em does not impose explicit constraints on the orders of all
  matching phrases} between the translated code and the reference
code, even though it enforces the order among the tokens in a $n$-gram 
matching phrase. 
%
For~example, given the reference, "\textbf{A B C D}", the BLEU scores
of the two translated results, $t_1$: "\textbf{A B} E \textbf{C D}",
$t_2$: "\textbf{C D} E \textbf{A B}" are the same, because the number
of matching phrases such as "{\bf A B}" and "{\bf C D}" in the two results is
similar. Therefore, for a given reference code and a translated
result, there are many possible variants that have the same BLEU
scores as the translated one. However, for migrated results,
the difference in the order of phrases between two results might lead
to the considerable semantic difference between them. For
instance, if "\textbf{A B}", "\textbf{E}", and "\textbf{C D}"
represent for the blocks of code for opening a file, reading the opened
file, and closing it, $t_1$ and $t_2$ are significantly
different in term of functionality. Hence, {\em by following that
  principle, the results from p-mppSMT achieve the BLEU score equivalent
  to the results of mppSMT, but possibly with a lower semantic~score}.

\subsection{Data Collection}

We collected a parallel corpus of 34,209 pairs of methods written in
both Java and C\texttt{\#}. Those methods were created manually by
developers, and used in 9 open-source systems that were originally
developed in Java and then ported to C\texttt{\#}.
%
Both Java and C\# versions have been used in practice and
research~\cite{ase15}.
Not all of the methods in Java version has a respective one in the
C\texttt{\#} version. To collect respective methods in each pair of
corresponding versions, we built a tool to conservatively search for
only the methods having the same signatures in the classes with the
same/similar names in the same/similar directory structures in both
versions. Such pairs of methods likely implement the same
functionality. We also manually verified a randomly selected
sample set to have high confidence that the method pairs are in fact
the respective ones. In total, we collected 34,209 respective methods
(Table~\ref{tab:systems}).


We used the chosen SMT-based migration tools described earlier on the
above dataset. We applied ten-fold cross validation by dividing all
aligned methods into ten folds with equal numbers of methods. To
test for a fold, we used the remaining folds for training. The
resulting methods were compared against the referenced ones in the
ground truth.

\subsection{Semantic Accuracy} 


%
Remind that our goal of the experiment is to (in)validate whether BLEU
evaluates effectively the translated results based on the semantic accuracy 
of the migrated code from the tools with regard to the manual-migrated code 
in the ground truth. 
{\em Semantic accuracy} between the result from an SMT-based migration
tool and the reference code from the ground truth is the similarity
between their respective functionality. 
If two methods perform similar operations on the respective given
inputs, they are semantically similar, even interchangeable. A pair of
methods can have the same functionality despite of their difference in
term of code structure and code elements.
For example, a method using a \code{for} loop and another using a
\code{while} loop, they still can perform the same functionality even
though their lexical representations are much different. 
%
%
To evaluate similar functionality between the translated code and the
original code, we adopted the same methodology as in the work by {\em
  Tsvetkov et al.}~\cite{tsvetkov-acl15} by using human judgment in
measuring semantic accuracy.
%
%
Specifically, we conducted a study with a human subject to manually
evaluate the migrated code from the SMT-based migration tools.


\emph{1. Sample Size}. 
%
With our dataset containing 34,209 pairs translated methods, we aimed
to achieve the confidence level of 95\% and the margin of error of
5\%. Thus, for~one~model, according to Wiley Statistics Online
Reference~\cite{geek_2015}, we~randomly sampled 375 pairs of methods
from the dataset for~evaluation.


\emph{2. Setting}. The human subject of our study is a developer who
is fluent and has more than 8 years of programming experience in both
Java and C\texttt{\#}. The evaluator was given each pair of methods in
C\texttt{\#}: one is the translated by a SMT tool and another one is
the reference code originally written by developers.
%
(S)he was also given the original Java code to understand the
requirements of the migration task for this method. Moreover, (s)he
was provided with the Github links to the corresponding projects from
which those methods were extracted (in both Java and C\texttt{\#}
versions). This would give him/her the context of the migrated
code.

\emph{3. Scoring.} The developer as the human subject was told to give
a score for each of 375 pairs of methods. The key criterion for
evaluating the migrated results is the semantic accuracy with regard
to the same functionality between the migrated code and the reference
code in C\texttt{\#}.

If (s)he recognizes the same functionality between them, a perfect,
highest score of 4 must be given to the totally correct result. If
(s)he finds that the migrated code and the original one do not perform
the same functionality, a score of lower than 4 must be given
(inaccuracy). In this case of inaccurate results, we could ask the
evaluator to give a score to indicate the degree of semantic
inaccuracy. If so, the evaluator needs to quantify how close the
migrated methods in C\texttt{\#} are with respect to the functionality
of the respective reference code.
It is not straightforward to provide a guideline for such
quantification of the functional similarities across several pairs of
resulting code and original code.
On the other hand, the more the translated code functionally similar
to the original code, the more likely the users are willing to use the
translated code as the starting point for their migration task. That
is, the willingness to reuse the migrated code reflects its degree of
translation accuracy.
%
%
Therefore, for the cases of inaccurate results, we chose to ask the
evaluator the questions relevant to the willingness to spend efforts
to reuse such results.

Particularly, if the migrated result was totally incorrect and (s)he
finds it totally useless and does not want to fix it, (s)he must give
a lowest score of 0 (\ie the migrated code is totally incorrect).
If (s)he is willing to fix the migrated code and reuse it, a score of
3 must be given (code seems to be not exactly correct, however, the human
subject is willing to fix it). In contrast, if (s)he finds the
migrated code incorrect, and is not willing to reuse it, (s)he must
give a score of 1 (code seems to be incorrect in which some parts
could be reused). Finally, if the human subject is undecidable on
whether to reuse the migrated code, (s)he must give a neutral score of
2.

With this scheme, we are able to evaluate the quality of the migrated
code with the integration of both the semantic accuracy of the
resulting code with regard to the reference code in the ground truth,
and the willingness to correct the translated code to achieve the same
functionality as the original~code.


%

In summary, the score range is as follows:

\begin{compactitem}

\item {\bf Four}: A score of 4 means that the pairs of methods are
  identical in term of functionality, and the translated method can be
  used as-is.

\item {\bf Three}: A score of 3 means that the translated code seems to be
  correct. Even though it needs some minor modifications, one is
  willing to reuse the result.

\item {\bf Two}: A score of 2 means that the human subject cannot
  decide whether to reuse the translated code or not.

\item {\bf One}: A score of 1 means that the translated code seems to be
  incorrect, and even though some parts of the result are reusable,
  one is {\em not} willing to fix the code.

\item {\bf Zero}: a score of 0 means that the translated code is totally
  incorrect and useless, and it is better to re-write it entirely
  rather than reuse the result.

\end{compactitem}




We conducted the same human study for the translated results from all above models with a total of 2,250 manual assignments of such scores. Let us call them \textbf{{\em
    semantic scores}}.

\section{Empirical Results on BLEU Scores}
\label{sec:bleuresult}


\subsection{Correlation between BLEU Scores and Semantic Scores}


To disprove {\bf Cond\_1} (BLEU score reflects well the semantic
accuracy of results translated by a particular model), we show the
relation between BLEU scores and human judgments via semantic scores.
We use Spearman's correlation coefficient~\cite{geek_2015} to gauge how
strong their relation is. The correlation coefficient has a value
between [-1, 1], where 1 indicates the strongest positive relation, -1
indicates the strongest negative relation, and 0 indicates no
relation.

\begin{figure}
\caption{BLEU Scores vs Semantic Scores (lpSMT Model)}
\centering
\includegraphics[width=2.9in]{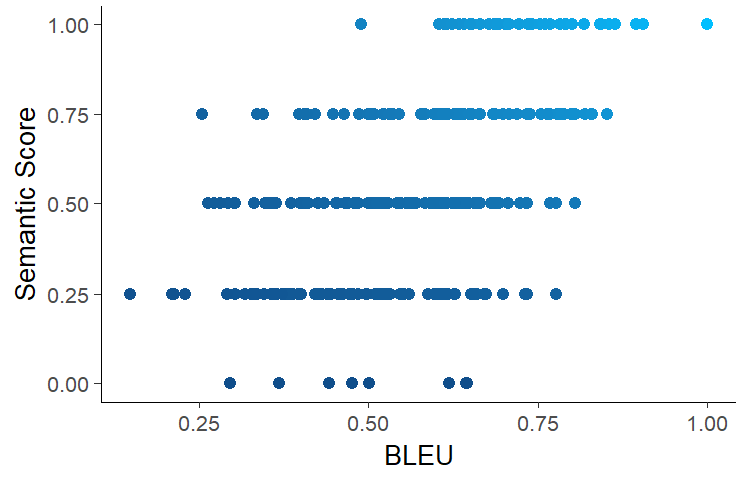}
\label{fig:BleuSemlpSMT}
\end{figure}

\begin{figure}
\caption{BLEU Scores vs Semantic Scores (mppSMT Model)}
\centering
\includegraphics[width=2.9in]{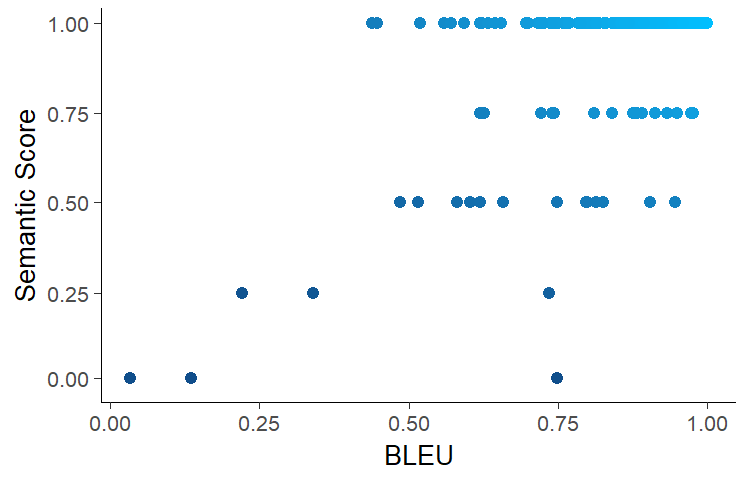}
\label{fig:BleuSemMppSMT}
\end{figure}

Fig.~\ref{fig:BleuSemlpSMT} and Fig.~\ref{fig:BleuSemMppSMT} show the
scatter plots between two metrics: BLEU and Semantic. Each point
represents the scores of a pair of methods where its $x$-axis value is
for BLEU scores and $y$-axis value is for semantic scores. The
correlation coefficient between BLEU and semantic scores for the model
mppSMT is 0.523 and for the model lpSMT is 0.570. These positive values
are closer to 0.5 than to 1.0. This means there is a {\em positive but weak
relation} between BLEU score and semantic score. The weak correlations 
between the metrics on the results translated by lpSMT and mppSMT are
demonstrated in Fig.~\ref{fig:BleuSemlpSMT} and Fig.~\ref{fig:BleuSemMppSMT}.



\subsubsection{{\bf Higher BLEU does not necessarily reflect higher~semantic accuracy}}

In Fig.~\ref{fig:BleuSemlpSMT}, for many specific values of BLEU, the
corresponding semantic scores spread out in a wide range. For
instance, for the BLEU score of 0.75, the corresponding semantic scores
are 0.25--1.
Thus, from this observation, we conclude that the results migrated by
these models with {\em high BLEU scores might not achieve high
  semantic~scores}.
%

There are two main reasons for this result in our dataset.  First, the
translated methods might have multiple correct phrases, but in an
incorrect order, those methods can be incorrect, even not compilable.
%
For example, in Fig.~\ref{fig:issueexample2}, the translated method
misplaces the position of '\{', making the method have a low
semantic score, however, it has high BLEU score.
%
In other cases, the migrated results are incomplete code missing the
elements that are trivial for the translation model, yet important
with respect to the syntactic rules of the target language. For
example, the result contains mostly keywords and separators such as
\code{if}, \code{public}, \code{()}, but misses out the important
program elements such as function calls or variable names. In this
case, it will have low semantic score while having a moderate to high
BLEU score. These cases indicate the limitation of BLEU metric
in evaluating the translated results in programming language in which
syntax rules are well-defined.


\subsubsection{{\bf Higher semantic accuracy is not necessarily reflected by
higher BLEU score}} In the results translated by mppSMT
(Fig.~\ref{fig:BleuSemMppSMT}), for a particular value of the semantic
score, there can be many corresponding BLEU values spreading out in
a~wide range. Specifically, with the semantic score of 1, the BLEU
scores can vary from 0.5 to 1.0. Thus, it can be concluded
that for this model, the migrated code achieving {\em higher semantic score does not necessarily have higher BLEU~score}.

From our data, we observe that there are two main reasons leading to
that result.
First, a translated method can use code structure different from the
reference one to perform the same
functionality. Fig.~\ref{fig:mppSMT_example} shows an example that got
a maximum semantic score, but has low BLEU score of 0.4. In this
example, the translated method uses a \code{for} loop instead of a
\code{foreach} loop as in the reference code. The second reason is the
whitespace issue. For example, in Fig.~\ref{fig:mppSMT_example}, the
translated code has the tokens \code{IsSimilar(}, but the reference
code has \code{IsSimilar (}. The former is interpreted as one token
while the latter is interpreted as two. This situation reduces
the precision on phrases, but the human subject still evaluated the
result with a high semantic score. By this experiment, we also
empirically verify that the focus on the lexical precision of BLEU
makes it unable to capture other code-related aspects such as program
dependencies that contribute to program semantics. This might lead
to the ineffectiveness of BLEU in reflecting the semantic accuracy of
translated results.

In brief, {\em BLEU does not reflect well the
semantics of source code, and it is not suitable for evaluating the 
semantic accuracy of the result from an SMT-based code migration model}.

\begin{figure}[t]
\centering
\begin{lstlisting}[basicstyle=\scriptsize\sffamily,language=Java]
		// C# reference code: 
		public bool IsSimilar ( String name ) {
			int idx = name. IndexOf ( '[' );
			for ( String n : part ) {
				if (n. StartsWith ( name ) ) {
					return true;
				}
			}
			return false;
		}
		// C# code translated by mppSMT:
		public bool IsSimilar( String name ) {
			int idx = name. IndexOf ( '[' );
			for ( int k = 0; k < part. Count ; ++k ) 
				if ( part [ k ] . StartsWith ( name ) ) {
				return true;
				}
			}
			return false;
		}
		// C# code translated by p-mppSMT
		public bool } String name ) {
			int idx = name. IndexOf ( '[' ); 
			for ( . ] k [ part ( if ) ++k ; 
			Count part. < k 0; = k int StartsWith ( name ) ) 
			{ return true; } } 
			return false; IsSimilar(
\end{lstlisting}
\caption{Translated Code and the Corresponding Reference Code}
\label{fig:mppSMT_example}
\end{figure}

\subsection{The Use of BLEU in Comparing SMT Migration Models}

To confirm the difference in the semantic scores for the two
models even though their results have the same BLEU~score, we performed the
paired $t$-test\cite{geek_2015} on 375 results translated by mppSMT and
p-mppSMT with the confidence level of
0.95 and a null hypothesis that {\em the semantic scores of two models
are equal}. With such setting for this data set, the $t$-critical value is 1.967,
and the confidence interval for the difference between the mean semantic scores
of two models is $\pm0.04$. It means the difference must be larger than 0.04 to
be considered significant. The $p$-value (the probability
that the results occurred by chance) of our test is 2.81E-85, which is much
smaller than significant level of 0.05. Thus, we rejected the null hypothesis and concluded
that the semantic scores of two models are different with the confidence
level of~95\%.

Additionally, in our experiment on the results translated by mppSMT
and p-mppSMT, the average BLEU score for these models are
identical. Meanwhile, the overall semantic score for p-mppSMT is 0.62,
which is lower than the average for mppSMT of 0.88. This means that
the deviant of BLEU in comparing the two models is nearly two
levels of semantic score lower, despite that they have identical BLEU
score.

Examining the translated code, we found that p-mppSMT produces the
code with lower semantic accuracy due to the swapping of the locations
of the expressions and/or statements in the code. That leads to the
completely different program semantics and even not compilable code. For
example, in Fig.~\ref{fig:mppSMT_example}, the phrase with the method
name \code{isSimilar} is placed at the end of line 27, making the code
incorrect. Therefore, the semantic score of such result is lower
than the result from mppSMT.

In our work, we additionally conducted another experiment in the same
manner on two pairs of SMT-based migration models: lpSMT~\cite{fse13}
and GNMT~\cite{gnmt}, and mppSMT~\cite{ase15} and
GNMT~\cite{gnmt}. The results from our $t$-tests allowed us to
conclude that an improvement in BLEU score is not sufficient nor
necessary to reflect a higher semantic score with the confidence of
0.95. Thus, the experimental results also invalidate the conditions. All
results can be found on our website~\cite{ruby-website}.

In brief, {\em BLEU is not effective in comparing SMT-based code
  migration models}. Finally, because both of the above necessary
conditions are false, we can conclude that {\em BLEU score does not
  measure well the quality of translated results}.


\section{Alternative Metrics}
\label{sec:alternatives}
%

As Section~\ref{sec:bleuresult} shows that BLEU is ineffective in
evaluating translated code, in this section, we will present our
evaluation of the effectiveness of various other metrics. The
effectiveness of a metric is expressed in its abilities to measure the
similarity between the reference code and the migrated code in term of
program semantics according to a programming~language.


In general, a programming language can be characterized by three
important parts: lexeme, syntax, and semantics, which correspond
with three important parts in a natural language: vocabulary, grammar,
and meaning.
These parts of a programming language could be represented respectively in three 
abstractions: tokens (text), abstract syntax tree (AST), and program 
dependence graph (PDG).
There are a number of clone detection approaches~\cite{clone-tse07},
which are token-based~\cite{ccfinder},
tree-based~\cite{baxter98,deckard}, and PDG-based~\cite{deckard2} to
measure the similarity of source code in these three representations.
These techniques have shown that the duplication of source code in
term of functionality can be detected more precisely when the higher
level representations are applied from text, AST, to
PDG~\cite{clone-tse07,deckard2}.
Based on that, we have the following research question: \textit{Does
  the metric that measures results in the higher abstraction level
  reflect better the similarity of the translated code and the
  reference code in term of program semantics?}

%
%
To answer the question, we conduct our experiments to evaluate the
effectiveness of three metrics: \textit{string similarity},
\textit{tree similarity}, and \textit{graph similarity} that measure
the results in the text, tree and graph representations for code
respectively, in reflecting the semantic accuracy of translated
results.

\subsection{Code Similarity Metrics}

We define the similarity scores on token-based, tree-based and
graph-based representations between the translated and the
reference code as the complement of string edit
distance~\cite{levenshtein}, tree edit distance~\cite{oopsla10}, and
graph edit distance~\cite{sanfeliu}, respectively.

\textit{String similarity (STS)} is the metric to compare the
similarity of source code that is represented as a sequence of
tokens. The similarity between the translated result $T$ and its
reference code $R$ is computed as:
$$STS(R, T) = 1 - \frac{SED(S_R, S_T)}{max\left(length(S_R), length(S_T)\right)}$$
where $SED(S_R, S_T)$ is the string edit distance between the
reference sequence $S_R$ and the translated sequence $S_T$. $SED$
measures efforts that a user must edit in term of the tokens that need
to be deleted/added to transform the resulting code into the correct one;
$length(t)$ refers to the length of the sequence~$t$.



\textit{Tree similarity (TRS)} is the metric used to measure
the similarity between two ASTs representing for the translated code $T$
and the reference code $R$:
$$TRS(R, T) = 1 -\frac{TED(AST_R, AST_T)}{size(AST_R) + size(AST_T)}$$ where $TED(AST_R, AST_T)$ is the edit
distance between the ASTs of the reference code $AST_R$ and the
translated code $AST_T$. $TED$ is calculated by the minimum number of
editing operations on AST nodes ({\em add}, {\em delete}, {\em
  replace}, and {\em move}) to sequentially make $AST_R$ and $AST_T$
identical \cite{oopsla10}; and $size(t)$ returns the number of
nodes in the AST $t$.
In this work, we define \textit{graph similarity (GRS)} between the
translated result $T$ and the reference code $R$ represented as PDGs
by using graph edit distance~\cite{sanfeliu}. That is:
$$GRS(R, T) = 1-\frac{GED(PDG_R, PDG_T)}{size(PDG_R)+ size(PDG_T)}$$ where $GED(PDG_R, PDG_T)$ is the edit distance
between the PDG of the reference code $PDG_R$ and the PDG of the
translated code $PDG_T$; $size(g)$ is the sum of the number of
vertexes and edges of the graph $g$. Specifically, $GED(PDG_R, PDG_T)$
is computed as the minimum number of graph edit operations to
transform one graph to another. The feasible graph edit operations on
vertexes and edges include {\em insert}, {\em delete}, and {\em
  substitute}.  However, computing the graph edit distance between two
graphs is NP-hard.  In this work, we use Exas~\cite{fase09}, a highly
precise approximate technique to compute $GED$.

\subsection{Experimental Results on Alternative Metrics}


To evaluate the abilities of STS, TRS and GRS in reflecting the
semantic accuracy of translated code, we conducted three experiments
for these three metrics in the same manner as the experiment described in
Section~\ref{sec:bleuresult}.

The correlation coefficient results between each of the three metrics
with the semantic score are displayed in
Table~\ref{table:correlation}. These results follow a similar trend:
for each model, the correlation coefficients increase 
accordingly to the abstraction levels of code from text, syntax, to semantic
representations in three metrics. For example, for mppSMT, the correlation
coefficient between STS and semantic score is 0.549, whereas that
for TRS is much greater at 0.820. The highest value is the
correlation coefficient between GRS and semantic score at 0.910. A
reason for this phenomenon is that for a pair of source code, if they
are textually identical, they have the same AST representation, and
the exact-match in AST leads to the equivalence between the
corresponding PDGs. Meanwhile, the equivalence in a higher
abstraction level does not necessarily mean the equivalence in a
lower abstraction. For example, Fig.~\ref{fig:mppSMT_example} shows
two methods with equivalent PDGs, but with textually~different. 


From these results, we conclude that the metric that measures the
results in the higher abstraction level is better in reflecting
semantics accuracy. Therefore, we propose a metric to evaluate the
quality of translated code, which uses the measurement of the
similarity of translated code and expected results in multiple
representations at different abstraction levels
(Section~\ref{sec:proposal}).




\begin{table}
\centering
\caption{Correlation between Metrics with Semantic Score}
\begin{tabular}{|c|c|c|c|c|}
\hline
 & STS & TRS & GRS\\
\hline
mppSMT  & 0.549 & 0.820 & 0.910 \\
\hline
lpSMT  & 0.533 & 0.786 & 0.823 \\
\hline
GNMT & 0.692 & 0.734 & 0.927 \\
\hline
\end{tabular}
\label{table:correlation}
\end{table}

\section{{\model}-An effective metric for code migration}
\label{sec:proposal}


\subsection{Design}

%

In this work, we introduce {\model}, a novel metric to evaluate
SMT-based code migration tools.  From the empirical results
listed in Section~\ref{sec:alternatives}, we can observe that the
metric measuring translated results in a higher level representation
can achieve higher correlation coefficient with semantic score.
However, the higher representation like PDG cannot always be
constructed due to the missing syntactic or semantic-related
information in the translated results that is required to construct
ASTs and PDGs. For example, the sets of methods that are translated by
lpSMT, and applicable for GRS and TRS are quite limited, 75 and 123
(out of 375 methods), respectively (the respective numbers for mppSMT
are 239 and 292, and those for GNMT are 128 and 155).
In brief, we cannot use GRS as a sole metric to evaluate the migrated code
because {\em not all resulting migrated code is sufficiently correct
  to build the PDGs or even~ASTs}.

To address this problem, we adopt an idea from machine learning,
called {\em ensemble methods}~\cite{ensemble}. Ensemble methods
use multiple learning models to obtain better
predictive performance than could be obtained from any of the
constituent learning models alone.
To apply to our problem, we would use the decision made by the most
reliable metric, which is GRS metric. In the case that the most
reliable metric is not available (due to the fact that a PDG cannot
be built from incorrectly migrated code), we would use the next
reliable one, which is TRS. TRS in turn would face the same issue if
the migrated code is not syntactically correct. In this case, we would
resolve to use STS as it is always computable for any given piece of
code with regard to the reference code. In this work, we design 
{\model} as an ensemble expert-based metric. Generally, {\model} is 
computed by the following formula:

$$
RUBY(R, T) = \begin{cases}
				GRS(R, T), 	& \mbox{if } PDGs\mbox{ are applicable} \\
				TRS(R, T), 	& \mbox{if } ASTs\mbox{ are applicable} \\
				STS(R, T), 	& \mbox{otherwise}
			\end{cases}
$$
where $R$ and $T$ are the reference and translated code respectively;
$GRS$, $TRS$, and $STS$ were defined in Section~\ref{sec:alternatives}.

\subsection{Empirical Results for {\model}}
\subsubsection{Correlation of {\model} and Semantic Scores}
To evaluate the effectiveness of {\model} in comparison to BLEU, we
conducted experiments in the same manner with BLEU on three models
GNMT, mppSMT, and lpSMT.

%

Our result shows that the correlation coefficients between {\model}
and semantic score are 0.764, 0.815, and 0.747 for the three models
GNMT, mppSMT, and lpSMT respectively. They are higher comparing to the
correlation of BLEU of 0.658, 0.523, and 0.570.
These values indicate a strong positive linear relationship between
two quantitative metrics. Thus, they show the effectiveness of
{\model} in reflecting semantic accuracy.



In general, the correlation coefficient between {\model} and semantic score cannot be 
stronger than GRS or even TRS. However, {\model} scores are available for any pair 
of the translated result and the reference code, while GRS and TRS are applicable only to 
a subset of translated results. 
%


\subsubsection{The Use of {\model} in Model Comparison}
We performed an evaluation on RUBY being used in comparing the two
models mppSMT and p-mppSMT.
%
Then, we also evaluated {\model} in comparing three real-world
SMT-based migration models lpSMT~\cite{fse13}, mppSMT~\cite{ase15},
and GNMT~\cite{gnmt}.  
%
For any two models $M$ and $M'$, we proceeded as follows. First, we
performed $t$-test (at the confidence level of 95\%) with the mean
values of semantic scores on the two sets of results migrated by the
two models. We then processed $t$-test in the same manner on the two
sets of {\model}'s scores for the results migrated by the two
models. If the decision whether $M$ is better than $M'$ by {\model} is
consistent with that by the semantic scores, we consider {\model} as
effective in comparing the models.


\noindent \textbf{Comparing mppSMT and p-mppSMT:} In this experiment, 
to compare the two models mppSMT and p-mppSMT, by using $t$-test, we
were able to reject the null hypothesis that \textit{the semantic
scores of two models are equal} with $p$-value of 2.81E-85. Because
the average semantic score of mppSMT is greater than p-mppSMT by
0.268, we conclude that the translation quality of mppSMT is better
than p-mppSMT based on the semantic scores.
To compare mppSMT and p-mppSMT using {\model}, we performed another
$t$-test with the null hypothesis that \textit{the {\model} scores of
two models are equal}. We were able to reject it with $p$-value of
7.78E-65. The confidence interval for the difference between means is
$\pm0.01$ at 95\% level of confidence. For mppSMT, the mean value of
{\model} scores is higher than that of p-mppSMT (0.258). Hence, the
conclusion that mppSMT is better than p-mppSMT is consistently stated
by {\model} and by the semantic scores.
Thus, the effectiveness of {\model} in comparing
those two models
is confirmed.

\noindent \textbf{Comparing real-world models:}
We also made pairwise comparison among the three models
lpSMT~\cite{fse13}, mppSMT~\cite{ase15}, and GNMT~\cite{gnmt}. In our
$t$-tests, for all pairs, we were able to reject the null hypotheses
that \textit{the semantic scores of two models are equal}
and \textit{the {\model} scores of two models are equal}. When
comparing the average scores, {\model}'s decision is always consistent
with the semantic scores.
The results including the $p$-values and the mean values for the
translated results of the models are shown in
Tables~\ref{table:tTestResult} and~\ref{table:avgRubySem}. 
As seen, for all pairs of models, an improvement in {\model} scores
when comparing the models always reflects an improvement in
translation quality (represented by semantic scores). In brief, {\em
{\model} can be used in comparison SMT-based code migration models
because it reflects well the semantic scores in such~comparison}.
%
\begin{table}
\centering
\tabcolsep 3pt
\caption{$p$-value Results in Model Comparison}
\begin{tabular}{|c|c|c|c|}
\hline
 & GNMT \& mppSMT & lpSMT \& mppSMT & lpSMT \& GNMT \\
\hline
Semantic  & 1.70E-35 & 3.32E-4 & 3.22E-26  \\
\hline
RUBY  & 2.27E-25 & 1.10E-3 & 9.72E-26  \\

\hline
\end{tabular}
\label{table:tTestResult}
\end{table}

\begin{table}
\centering
\caption{Mean Semantic Scores and {\model} Scores}
\begin{tabular}{|c|c|c|c|}
\hline
 & GNMT & lpSMT & mppSMT \\
\hline
Semantic & 0.714 & 0.879 & 0.914  \\
\hline
{\model} & 0.743 & 0.886 & 0.905  \\
\hline
\end{tabular}
\label{table:avgRubySem}
\end{table}


\subsection{Further Study}

\noindent {\em Experimental procedure:} To empirically investigate
further the characteristics of {\model} in reflecting the semantic
accuracy in migrated results, we collect the subsets of the results in
the set of all translation results that {\model} scores reflect well
and not so well the translation quality. The subset of results in
which {\model} reflects well the translation quality is the set with
the {\model} scores that correlate well with the semantic scores.
Therefore, we aim to collect in our sample dataset the largest subset
of the results whose {\model} scores correlate highest to the
semantic scores. The subset of results in which {\model} does reflect
well their translation quality, contains the remaining translation
results in the sample dataset.


%


To do that, we make use of Random Sample Consensus (RANSAC)
algorithm~\cite{Fischler:1981:RSC:358669.358692} to select two subsets
of the sample dataset of the results: 1) the largest subset of the
results that have highest correlation coefficient between {\model}
scores and semantic scores (Subset 1); and 2) the remaining
translation results in the sample dataset, in which they have
inconsistencies between semantic scores and {\model} scores (Subset 2).

RANSAC algorithm at first selects randomly the minimum number of
results to determine the model parameters. Therefore, the chosen
subsets depend on randomization. In order to achieve the Subsets 1 and
2, we independently run RANSAC 10 times on the sample dataset of
translation results.
The size of each subset for each time ranges from 224 to 315, while their correlation values are at the high rate (between 0.93 and 0.95).
In our experiment, we chose to investigate the
subset with the median of the correlation, 0.955, with the
size of 308.

\noindent \textbf{Subset 1:} On this subset, the code features related
to program dependencies and syntaxes that contribute to the semantics
of the translated methods are captured by {\model}. By manually
analyzing Subset 1, we categorize them into the following~cases.

First, there are 32 out of 308 cases that {\model} is able to capture
program elements having the same functionality but with the lexical
tokens different from the reference code. This feature is only fully
captured by using PDGs in {\model}. For example, in
Fig.~\ref{fig:mppSMT_example}, the translated code uses a \code{for}
loop, which is the same meaning as \code{foreach} in the reference
code. Moreover, access to an array element could be implemented in
different ways such as via \code{part[k]} or \code{Iterator}.

Additionally, there are 66 translation results that correctly contain
the key program elements such as API elements, method calls, and
variables as in the reference code, even though their usages are
incorrect. By comparing tree representations with TRS, {\model} can
capture these elements based on the common subtrees of translated and
reference code. In function \code{TestImaginaryTokenNoCopyFromTokenSetText} of
Fig.~\ref{fig:samples}, the variables \code{grammar} and
\code{found}, and the method call \code{Assert.AreEqual} appear in
both translated and reference code, but the first parameter is not correct.
%

Finally, in Subset 1, there are the translation results containing the
same structures and data flows as in the reference code, but lexically
differ from one another. For example, in Fig.~\ref{fig:samples}, the
structure of the method \code{CurveTo} in the translated version is
similar to the reference one with the condition branches \code{if} and
throwing exception \code{IllegalPdfSyntaxException}. {\model} reflects
this similarity via the similarity in the data and control
dependencies among program entities.

\begin{figure}[t]
	\centering
\begin{lstlisting}[basicstyle=\scriptsize\sffamily,language=Java]
		// Reference code
		public void TestImaginaryTokenNoCopyFromTokenSetText() {
			string grammar = "grammar_T;\n" + "options_{output=AST;}\n" + ...
			string found = execParser("T.g", grammar, "TParser", "TLexer", ...;
			Assert.AreEqual("(block_a_b_c)" + NewLine, found);
		}
		// Translated code
		public void TestImaginaryTokenNoCopyFromTokenSetText() {
			string grammar = "grammar_T;\n" + ...;
			string found = execParser("T.g", grammar, ...;
			Assert.AreEqual( < unk > + NewLine, found );
		}
		// Reference code
		public void CurveTo ( float x1 , float y1 ...) {
			if (inText) {
				if (writer.IsTagged()) {
					EndText();
				} else {
					throw new IllegalPdfSyntaxException(...);
				}
			}
			content.Append(x1).Append(' ').Append(y1)...;
		}
		// Translated code
		public void CurveTo ( float x1 , float y1 ...){
			if (inText) {
				if (writer.IsTagged()) {
					EndText();
				} else {
					throw new IllegalPdfSyntaxException(...).Append(font.).Append(...
					Append( < unk > ) ...
				}
			}
		}
\end{lstlisting}
	\caption{Examples that {\model} Reflects Well Semantic Accuracy}
	\label{fig:samples}
\end{figure}

\noindent \textbf{Subset 2:}
{\model} also suffers limitations. {\model} aims to measure the
similarity between the translated code and the reference one in the
highest applicable representation level among text, AST, and
PDG. However, in our sample set, there are some results that are
syntactically incorrect but still meaningful in human perspective. In
these cases, the effectiveness level of {\model} is only achieved at
the lexical level via STS. Thus, it is not as effective.

\subsection{Threats to Validity}

The dataset of 9 subject projects with +34K method pairs that we used
in our experiments might not be representative. However, it has been
used to evaluate SMT-based code migration models in existing
work~\cite{fse13,ase15,icsme16}. For {\model}, it is impossible to
perform the experiments on all migration models from any programming
language to another. We mitigate this threat by using it on the most
popular SMT-based migration models and from Java to C\texttt{\#}. Our
semantic scores were built by a human subject, which might make
mistakes. Our statistical validation methods might have statistical
errors.

\section{Related Work}

There exist many studies aiming to measure the functionality
similarity of source code, which utilize the similarities of
structures and 
dependencies~\cite{clone-tse07,roy09,baker97,ccfinder,cpminer,deckard,deckard2,horwitz01,baxter98}.
Specifically, several approaches in the {\em code clone detection}
literature have been proposed~\cite{clone-tse07,roy09}. Generally,
they can be classified based on their code representations. The
typical categories are {\em text-based}~\cite{ducasse-icsm99}, {\em
  token-based}~\cite{baker97,ccfinder,cpminer,mende08}, {\em
  tree-based}~\cite{baxter98,deckard}, and
{\em graph-based}~\cite{deckard2,horwitz01,liu06}.

The text-based~\cite{ducasse-icsm99} and token-based~\cite{ccfinder}
approaches are efficient, scalable, and independent of the programming
languages. However, they cannot detect the code clones with different
syntactic structures. AST-based approaches~\cite{baxter98} overcome
that limitation but is language-dependent and has higher computational
complexity. Other approaches~\cite{fase09,deckard} make use of vector
computation to represent tree-based structures to reduce such
complexity.  Deckard~\cite{deckard} introduced the use of vectors in
clone detection. In Exas~\cite{fase09}, the authors showed that vector
representation for tree-based fragments is more generalized. Deckard
counts only distinct AST node types in a subtree for a fragment, while
Exas captures structural features via paths and sibling sets.
%
NICAD~\cite{nicad08} detects near-miss clones using flexible
pretty-printing and code~normalization.

Graph-based clone detection approaches, though providing clones of the
higher level of abstraction, are of high complexity in detecting
similar subgraphs. Krinke's tool detects code clones via a program
dependence graph (PDG)~\cite{krinke01}. It finds the maximal
isomorphic subgraphs in a PDG by calculating the maximal $k$-limited
path induced subgraphs. Such induced subgraphs are defined as the
maximal similar subgraphs that are induced by $k$-limited paths
starting at two vertices. Their approach is more heavy-weight than the
use of vector-based calculation after structural feature extraction in
Exas.

Our study
is similar in nature to Callison-Burch {\em et al.}~\cite{Callison}'s
in NLP area. There exists criticism on BLEU as the authors argued that
an improvement in BLEU metric is not sufficient nor necessary to show
an improvement in translation quality.


The statistical NLP approaches have been applied in software
engineering. Hindle {\em et al.}~\cite{natural} use
$n$-gram~\cite{manning99} with lexical tokens to show that source code
has high repetitiveness. 
%
Raychev {\em et al.}~\cite{ethz-pldi14} capture common sequences of
API calls with $n$-gram to predict next API call.
NATURALIZE~\cite{barr-codeconvention-fse14} is a $n$-gram-based
statistical model that learns coding conventions to suggest natural
identifier names and formatting conventions.  $n$-gram is also used to
find code templates for a task~\cite{jacob10}, for large-scale
code mining~\cite{sutton-msr13}, for model
testing~\cite{tonella-icse14}, etc.

Tu {\em et al.}~\cite{tu-fse14} improve $n$-gram model with caching
capability for recently seen tokens to increase next-token prediction
accuracy.
SLAMC~\cite{fse13} associates to code tokens the semantic
annotations including their token and data types.
Maddison and Tarlow~\cite{tarlow14} use probabilistic context free
grammars and neuro-probabilistic language models to build language
model for source code. Their model is generative and does not use
semantic context. Oda {\em et al.}~\cite{hide-ase15} uses SMT to
generate English and Japanese pseudo-code from source~code.

\section{Conclusion}
In this work, we invalidated the use of BLEU for code migration. We
showed counterexamples to prove that BLEU is not effective in
reflecting the semantic accuracy of translated code and in comparing
different SMT-based code migration systems. Our empirical study
illustrated that BLEU has weak correlation with human judgment in the
task of measuring semantics of translated source code. Also, BLEU's
decision on the difference between models' translation quality is
inconsistent with that from semantic scores. We concluded that BLEU is
not suitable for source code migration, and proposed a new alternative
metric {\model} to replace BLEU. {\model} is a novel ensemble metric
that takes into account three representation levels of source code. We
validated {\model} on multiple SMT-based code migration models to show
its reliability in estimating semantic accuracy and comparing
translation quality between those models. {\model} is always definable
for any code fragment even when AST or PDG representations cannot be
built. With its advantages, {\model} can be used to evaluate
SMT-based code migration~models.

\section*{Acknowledgment}
This work was supported in part by the US National Science
Foundation (NSF) grants CCF-1723215, CCF-1723432, TWC-1723198,
CCF-1518897, and CNS-1513263.

\newpage

\balance

\bibliographystyle{IEEEtran}

\bibliography{ase18}

\end{document}